\documentclass[manuscript]{aastex}
\input{colordvi.tex}
\usepackage{natbib}
\usepackage{apjfonts}
\usepackage{amsmath}
\usepackage{graphicx}

\shorttitle{Pair Echo from High-z GRBs}
\shortauthors{Takahashi et al.}

\begin{document}
\title{Probing Early Cosmic Magnetic Fields through Pair Echos from High-Redshift GRBs}
%Detectability of Pair Echo from High-z GRBs
\author{
Keitaro Takahashi\altaffilmark{1},
Susumu Inoue\altaffilmark{2},
Kiyotomo Ichiki\altaffilmark{1} and
Takashi Nakamura\altaffilmark{2}
}

\email{keitaro@a.phys.nagoya-u.ac.jp}

\altaffiltext{1}{Department of Physics and Astrophysics,
Nagoya University ,Chikusa-ku Nagoya 464-8602 Japan}

\altaffiltext{2}{Department of Physics, Kyoto University,
Oiwake-cho, Kitashirakawa, Sakyo-ku, Kyoto 606-8502 Japan}

\begin{abstract}
We discuss the expected properties of pair echo emission
from gamma-ray bursts (GRBs) at high redshifts ($z \gtrsim 5$),
%\Red{[Note change]},
their detectability, and the consequent implications for
probing intergalactic magnetic fields (IGMFs) at early epochs.
Pair echos comprise inverse Compton emission by secondary electron-positron pairs
produced via interactions between primary gamma-rays from the GRB
and low-energy photons of the diffuse intergalactic radiation, arriving with a time delay
%relative to the primary emission
that depends on the nature of the intervening IGMFs.
At sufficiently high $z$, %$z \gtrsim 5$,
the IGMFs are unlikely to have been significantly contaminated by astrophysical outflows,
and the relevant intergalactic radiation %at rest-frame infrared wavelengths
may be dominated by the well-understood cosmic microwave background (CMB).
Pair echoes from luminous GRBs at $z \sim 5-10$
may be detectable by future facilities such as the Cherenkov Telescope Array
or the Advanced Gamma-ray Imaging System, %or 5@5 array,
as long as the GRB primary emission extends to multi-TeV energies,
the comoving IGMFs at these redshifts %$z \sim 10$
are $B \sim 10^{-16}-10^{-15}~{\rm Gauss}$,
%\Red{[Note change]}
and the non-CMB component of the diffuse intergalactic radiation %at low $z$ %for $z < 10$
is relatively low.
%Although observationally challenging,
Observations of pair echos from high-$z$ GRBs
can provide a unique, in-situ probe of weak IGMFs during the epochs
of early structure formation and cosmic reionization.
\end{abstract}


%%%%%%%%%%%%%%%%%%%%%%%%%%%%%%%%%%%%%%%%%%%%%%%%%%%%%%%%%%%%%%%%%%%
\section{Introduction \label{sec:intro}}
%%%%%%%%%%%%%%%%%%%%%%%%%%%%%%%%%%%%%%%%%%%%%%%%%%%%%%%%%%%%%%%%%%%

Extragalactic sources of high-energy gamma-rays such as gamma-ray bursts (GRBs) or blazars
can give rise to delayed secondary emission components known as pair echos
\citep{1995Natur.374..430P}.
%If the energies of primary photons from such objects
%extend to the GeV-TeV range,
Primary GeV-TeV photons from such objects can interact with
infrared (IR) to ultraviolet (UV) photons of the diffuse intergalactic radiation
to create electron-positron pairs relatively far away from the source, %GRB itself
%as long as there is sufficient opacity to the process.
%These charged particles
which can then be deflected by intergalactic magnetic fields (IGMFs)
before emitting secondary gamma-rays via inverse Compton (IC) upscattering
of mainly cosmic microwave background (CMB) photons,
reaching the observer with a characteristic time delay relative to the primary photons. 
This pair echo emission depends on the properties of the intervening IGMFs
and hence constitute a valuable probe of their nature
\citep[e.g.][]{1995Natur.374..430P,2002ApJ...580L...7D,
2004ApJ...613.1072R,
2008ApJ...682..127I,
2008ApJ...687L...5T,2008ApJ...686L..67M,2009MNRAS.396.1825M}.
%2002ApJ...580.1013D,2004ApJ...604..306W,2004MNRAS.354..414A,2007ApJ...671.1886M,2007ApJ...656..306C
%\Red{[Removed references not discussing IGMF explicitly. Include Neronov \& Vovk 2010?]}
Depending on the IGMF,
the secondary emission can also result in a spatially-extended pair halo
around the primary gamma-ray source
\citep{1994ApJ...423L...5A,Neronov:2007zz,Dolag:2009iv,Elyiv:2009bx,Neronov:2009gh}.
%\Red{[Don't forget Aharonian et al. 1994]}
Very recently, using data from the {\it Fermi} Gamma-ray Space Telescope,
the existence of IGMFs of order $\sim 10^{-15}~{\rm Gauss}$ has been suggested
based on upper limits to the secondary emission for a few blazars \citep{2010Sci...328...73N}
or by the apparent detection of pair halos in stacked images of a large number of sources
(\citealp{2010arXiv1005.1924A}, see however, \citealp{2010arXiv1006.0164N}).

To date, numerous different kinds of physical scenarios have been proposed for the origin of IGMFs,
particularly in relation to processes in the early universe:
generation during cosmic inflation \citep{1988PhRvD..37.2743T,1992ApJ...391L...1R,
2007JCAP...02...30B,2007PhRvD..75h3516B} or other phase transitions \citep{1997PhRvD..55.4582S,2008PhRvL.101q1302C},
from cosmological perturbations around the cosmic recombination epoch
%\Red{[This description OK?]}
\citep{2005PhRvD..71d3502M,2005PhRvL..95l1301T,
2008PhRvD..77l4028T,2006Sci...311..827I,2009CQGra..26m5014M},
at ionization fronts \citep{2000ApJ...539..505G,2003PhRvD..67d3505L,2005A&A...443..367L,2010ApJ...716.1566A}
or shocks during cosmic reionization \citep{2005ApJ...633..941H,2010arXiv1001.2011M}, and
during nonlinear phases of large-scale structure formation \citep{1997ApJ...480..481K}.
%\Red{[Order changed to follow cosmic time sequence, OK?]}
%first stars \citep{2005ApJ...633..941H,2008ApJ...688L..57X},
Such studies were motivated by dynamo theories for the origin of galactic magnetic fields,
whereby weak, "seed" IGMFs existing before the formation of galaxies
can be amplified up to the observed levels during their evolution
\citep{2002RvMP...74..775W}.

%A related question concerns the relevance of magnetic fields to the formation of the first stars in the universe,
%which has been debated in the literature
%\citep[e.g.][]{2004ApJ...603..401T,2006MNRAS.371..444S,2007PASJ...59..787M,2008ApJ...688L..57X,
%2008ApJ...685..690M,2009ApJ...703.1096S}.
%2004ApJ...609..467M,2006ApJ...647L...1M

%\Red{[Paragraph to be completed.]}

%\Red{[Following paragraphs extensively modified.]}

Observational determination or constraints on IGMFs from such early epochs
would be crucial for understanding the origin of cosmic magnetic fields in general,
and may also give us new insight into the physics and astrophysics of the early universe.
However, a key concern is the possibility that other astrophysical sources of magnetic fields
such as supernova-driven galactic winds or quasar outflows
pollute the intergalactic medium (IGM) at later times and eventually dominate its magnetization.
Theoretical models of such effects \citep{2001ApJ...556..619F,2006MNRAS.370..319B} have suggested
that even at the current epoch, IGMFs in the central regions of intergalactic voids
remain uncontaminated and retain their original properties from high redshift
(save for the adiabatic effects of cosmic expansion),
so that pair echos and halos from low-redshift blazars or GRBs may still be a useful probe of early IGMFs.
Nevertheless, whether this is actually the case remains to be seen.
%Thus, it would not be an easy task to extract information on the generation mechanism
%from future measurements or limits on the amplitude and
%coherence length of intergalactic magnetic fields.

Thus, it would be highly desirable to have some means to probe IGMFs directly in-situ at sufficiently high redshifts,
before they are substantially affected by magnetized astrophysical outflows.
To this end, we focus on pair echos associated with high-redshift GRBs occurring at $z \gtrsim 5$.
%with redshift $z \gtrsim 10$,
GRBs have already been observed at such redshifts \citep[e.g.][]{2006Natur.440..184K},
at least up to $z \sim 8.2$ \citep{2009Natur.461.1258S,2009Natur.461.1254T},
and are expected at even higher $z$,
perhaps out to the earliest epochs of star formation in the universe \citep[][and references therein]{2007AIPC..937..532B}.
%associated with massive star formation
Moreover, they are established sources of luminous GeV gamma-ray emission
\citep[e.g.][]{1994Natur.372..652H,2009Sci...323.1688A}.
%with isotropic-equivalent luminosities reaching
Since the majority of GRBs so far do not show clear evidence of high-energy spectral cutoffs in the GeV region
\citep{2010arXiv1003.2452G},
it is not implausible that the spectra of at least some bursts extend to multi-TeV energies.
At $z \gtrsim 5$, the diffuse intergalactic radiation %at restframe IR to UV wavelengths
originating from stars and other astrophysical objects is quite uncertain %quasars
\citep[e.g.][]{2009MNRAS.399.1694G,2010MNRAS.404.1938I}.
However, depending on the cosmic star formation rate and other factors, its intensity may be low enough
(Y. Inoue et al., in preparation) so that
1) absorption of the primary GRB emission occurs mainly via $\gamma\gamma$ pair production
with the well-understood CMB, and
2) further absorption of the secondary pair echo emission is not severe.
The former point is crucial as it not only allows relatively reliable evaluations of the pair echo flux,
but also constraints on stronger IGMFs than compared to low-$z$ pair echos \cite[e.g][]{2008ApJ...687L...5T}
by virtue of the shorter length and time scales involved.
In this paper, we first discuss the basic physics
of pair echo emission at high-redshifts in \S \ref{sec:hizecho}.
Our numerical results are presented in \S \ref{sec:results},
followed by a discussion and summary in
\S \ref{sec:discussion} and \S \ref{sec:summary}, respectively.

%%%%%%%%%%%%%%%%%%%%%%%%%%%%%%%%%%%%%%%%%%%%%%%%%%%%%%%%%%%%%%%%%%%
\section{Pair echo emission at high redshifts \label{sec:hizecho}}
%%%%%%%%%%%%%%%%%%%%%%%%%%%%%%%%%%%%%%%%%%%%%%%%%%%%%%%%%%%%%%%%%%%

\subsection{Absorption of primary and secondary gamma-rays \label{sec:gammaabs}}

%\Red{[Discussion of EBL moved from \S 2.3 to here; first we must justify neglect of EBL at high-z.]}

Previous studies of pair echo emission have been limited to $z \lesssim 5$,
where primary GeV-TeV gamma-rays from sources such as blazars or GRBs
initially undergo $\gamma\gamma$ interactions with IR-UV photons of
the extragalactic background light (EBL),
%\Red{[Changed "CIB" to "EBL" everywhere. Beware that CMB itself enters IR at high-z.]},
mainly composed of the integrated stellar and dust emission from galaxies in this $z$ range
\citep[][and references therein]{2008AIPC.1085...71P}.
Although various theoretical models have been proposed for the EBL,
its detailed properties are still not known very accurately
as it is difficult to measure directly.
In recent years, important indirect constraints have been obtained from
searches for $\gamma\gamma$ absorption features in various GeV-TeV sources
by Cherenkov telescopes as well as the Fermi satellite,
all pointing to a low-$z$ EBL that is not far above the lower bounds derived from direct galaxy counts
\citep{2006Natur.440.1018A, 2008Sci...320.1752M,2010arXiv1005.0996F}.
%2007A&A...475L...9A,

At $z \gtrsim 5$, the situation is much more uncertain 
since the relevant observational information becomes very scarce,
let alone the lack of $\gamma\gamma$ constraints.
This is particularly true for $z \gtrsim 10$ where the only secure data is
the {\it WMAP} determination of the Thomson scattering optical depth.
Nevertheless, this epoch is currently of great interest for observational cosmology,
as it should encompass the formation of the first stars and galaxies in the universe,
as well as the reionization of the IGM after cosmic recombination \citep[e.g.][and references therein]{2007RPPh...70..627B}.
The only detailed discussion to date of $\gamma\gamma$ absorption in this cosmic reionization era
is the recent study by \citet[][hereafter I10]{2010MNRAS.404.1938I}, 
who employed semi-analytical models of cosmic star formation at $z=5-20$
including both Population II and III stars,
%that include both metal-poor, Population II stars and metal-free, Population III stars,
and which are consistent with a wide variety of existing high-$z$ observations
such as quasar Gunn-Peterson measurements, {\it WMAP} Thomson depth constraints, near-IR source count limits, etc.
According to their fiducial model of the high-$z$ EBL
\footnote{Here we adhere to the terminology of "EBL"
for referring to diffuse intergalactic radiation of astrophysical origin,
although strictly speaking, the term "background" is inappropriate
for UV intergalactic radiation in the cosmic reionization era,
which becomes increasing inhomogeneous at higher $z$.}, %diffuse intergalactic radiation,
appreciable attenuation can be expected
above $\sim$12 GeV at $z \sim 5$, down to $\sim 6-8$ GeV at $z \gtrsim 8-10$.

%%%%%%%%%%%%%%%%%%%%%%%%%%%%%%%%%%%%%%%%%%%%%%%%%%%%%%%%%%%%%%%%%%%
\begin{figure}[t]
\begin{center}
\includegraphics[width=12cm]{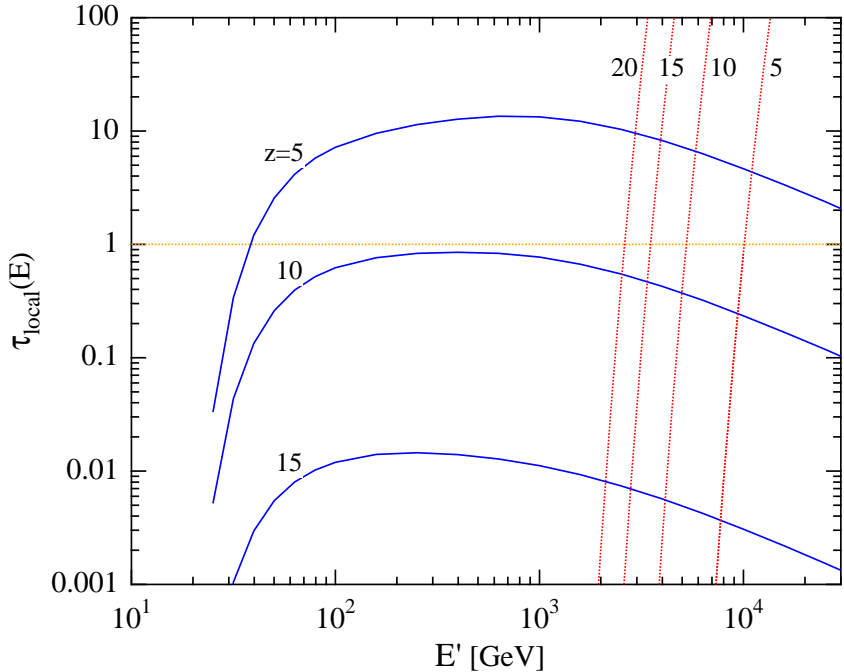}
\end{center}
\caption{
Local $\gamma\gamma$ optical depth $\tau_{\rm local}$ vs. rest-frame gamma-ray energy $E'_\gamma$
at redshifts $z$ as labelled for the fiducial EBL model of \citet{2010MNRAS.404.1938I},
compared with the contribution from the CMB.
%\Red{[Figure to be modified.]}
\label{fig:tauloc}}
\end{figure}
%%%%%%%%%%%%%%%%%%%%%%%%%%%%%%%%%%%%%%%%%%%%%%%%%%%%%%%%%%%%%%%%%%%%

Fig. \ref{fig:tauloc} shows estimates of the "local $\gamma\gamma$ optical depth" $\tau_{\rm local}$,
i.e. the optical depth across a Hubble radius at each $z$,
in terms of the rest-frame gamma-ray energy $E'_\gamma$
%versus rest-frame gamma-ray energy $E_{\rm rest}$,
for the fiducial model of I10. %\cite{2010MNRAS.404.1938I}.
%$l_H(z)=c/H_0 (\Omega_m (1+z)^3+\Omega_\Lambda)^{1/2}$,
While $\tau_{\rm local}$ is significant for $z \lesssim 10$
around $E'_\gamma \sim 10^2-10^4$ GeV,
that for $z \gtrsim 10$ becomes quite small,
owing to the declining star formation rate and hence the EBL intensity at higher $z$,
together with the reduced path length.
This is to be contrasted with the $\gamma\gamma$ opacity contribution from the CMB, also plotted in Fig.\ref{fig:tauloc},
which becomes increasingly prominent and moves to lower $E'_\gamma$ for higher $z$,
its evolution being governed simply by cosmic expansion.
At $z \gtrsim 10$, it is apparent that $\tau_{\rm local} \gtrsim 1$ at $E'_\gamma \gtrsim 3-6$ TeV
solely due to the CMB, whose Wien tail intrudes into the rest-frame IR band.

%\cite{2010MNRAS.404.1938I}
I10 also investigated some other models within their framework
that fit the current high-$z$ observations nearly equally well,
and found that they generally do not lead to large differences in the $\gamma\gamma$ opacity.
Nevertheless, it must be cautioned that by relaxing some of their basic assumptions,
e.g. regarding the stellar initial mass function or the quasar contribution,
a wider range of possibilities may very well be possible.
In fact, alternative models in which the star formation rates and EBL intensities at $z \sim 5-10$
are lower than I10 by as much as an order of magnitude, %\cite{2010MNRAS.404.1938I}.
close to the lower limits from deep near-IR counts \citep{2008ApJ...686..230B,2009arXiv0912.4263B},
may still be consistent with the available observations, as long as an appreciable Pop III component is included
at $z \gtrsim 10$ (Y. Inoue et al., in preparation).
Considering these uncertainties and limitations,
%amount of constraining observational data
%as well as in the theoretical understanding of the relevant astrophysics.
in addition to I10 that we refer to as the "high-EBL" case,
we also consider a "low-EBL" case for $z > 5$
where the EBL intensity is simply scaled down by a factor of 10 from I10. %\cite{2010MNRAS.404.1938I}
In the latter case, the CMB can dominate the $\gamma\gamma$ opacity for all redshifts above $z \sim 5$,
as is apparent in Fig. \ref{fig:tauloc}.

Below, we will apply these considerations not only to the initial absorption
of the primary gamma-rays, but also to further absorption of the secondary pair echo gamma-rays
as they propagate from $z \gtrsim 5$.
In view of the recent observational developments mentioned above,
for the EBL at $z < 5$, we adopt the "best fit" model of \citet{2004A&A...413..807K}
scaled by 0.5, which is a fair approximation to the current lower bounds on the EBL at $z=0$
as described in \cite{2010A&A...515A..19K}.
%\Red{[Better to use explicitly the public model of Kneiske \& Dole.
%But the resulting difference is probably not large, so OK with this for now.]}
%\citep{2008AIPC.1085..620K}

\begin{figure}[t]
\begin{center}
\includegraphics[width=12cm]{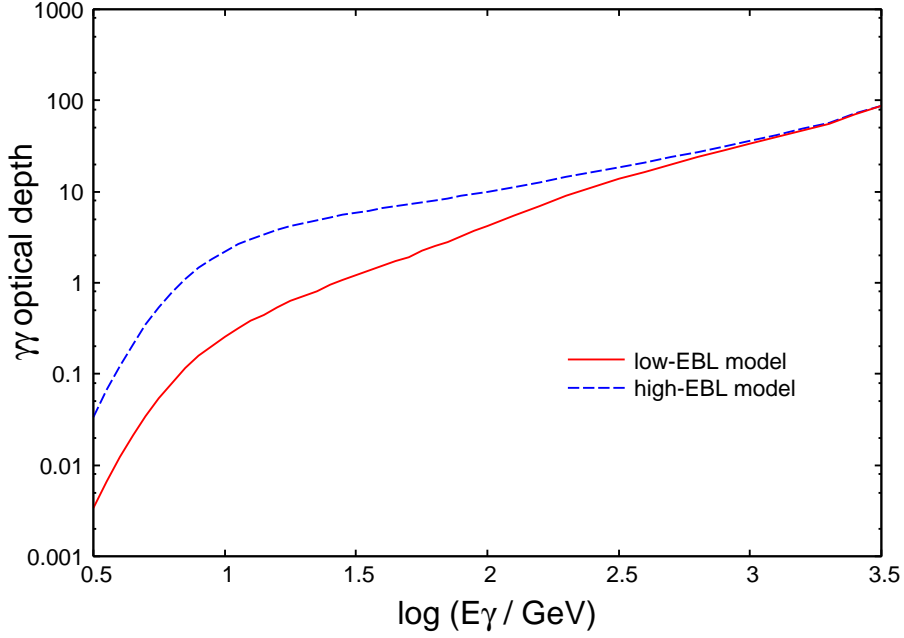}
\end{center}
\caption{$\gamma\gamma$ optical depth for a source at $z=10$ vs. observer gamma-ray energy $E_\gamma$ at $z=0$
for the low-EBL and high-EBL cases.
%\Red{[Please change "CIB" to "EBL" in figure. Better to have "$\gamma\gamma$" before "optical depth" in Y-axis label.]}
\label{fig:tau_comp}}
\end{figure}

The $\gamma\gamma$ optical depth for a source at $z=10$ observed at $z=0$
are compared for our low-EBL and high-EBL cases in Fig. \ref{fig:tau_comp}. 
The opacity at observer gamma-ray energy $E_\gamma \gtrsim 300~{\rm GeV}$
%\Red{[Must clarify "observer"!]}
is mostly due to the low-$z$ EBL and can be considered reasonably reliable.
On the other hand, that for lower energies $E_\gamma \lesssim 300~{\rm GeV}$ is caused by the high-$z$ EBL,
which is highly uncertain but strongly affects the observability of high-$z$ pair echos, as discussed below.
%($10~{\rm GeV} \lesssim E_\gamma \lesssim 300~{\rm GeV}$)

%%%%%%%%%%%%%%%%%%%%%%%%%%%%%%%%%%%%%%%%%%%%%%%%%%%%%%%%%%%%%%%%%%%
%%%%%%%%%%%%%%%%%%%%%%%%%%%%%%%%%%%%%%%%%%%%%%%%%%%%%%%%%%%%%%%%%%%

\subsection{CMB-triggered pair echos  \label{sec:cmbecho}}

%original discussion \citep{1966PhRvL..16..252G}

Following the above discussion, we proceed under the assumption that the only radiation field
responsible for the initial $\gamma\gamma$ interaction is the CMB.
This would be valid for all redshifts $z \gtrsim 5$ in the low-EBL case,
but only for $z \gtrsim 10$ in the high-EBL case.
We begin by outlining the basic phenomenology of CMB-triggered pair echos
(for more details on the general physics of pair echos, see \citet{2008ApJ...682..127I}).
Quantities such as photon energy as measured in the cosmological rest frame at redshift $z$
are designated with primes, whenever distinction is required between that observed at $z=0$,
unless otherwise noted.
%\Red{[Make clearer distinction between $z=0$ observer frame and rest-frame.]}

The characteristic photon energy and number density of the CMB around its spectral peak are respectively
\begin{equation}
\epsilon'_{\rm CMB,pk}
\approx 2.4 \times 10^{-3}
        \left( \frac{1+z}{10} \right) {\rm eV},
~~~
n_{\rm CMB,pk}
\approx 4.1 \times 10^5
        \left( \frac{1+z}{10} \right)^3 {\rm cm}^{-3}.
\end{equation}
%\Red{[Must clarify "peak"! Changed $T_{\rm CMB}$ to $\epsilon'_{\rm CMB,pk}$.]}
The typical energy of gamma-rays that can produce pairs with these CMB peak photons is
\begin{equation}
E'_{\gamma,pk}
= \frac{m_e^2}{2 \epsilon'_{\rm CMB,pk}}
\approx 54 \left( \frac{1+z}{10} \right)^{-1} {\rm TeV},
\end{equation}
where $m_e$ is the electron mass, and we choose units with $c=1$.
The $\gamma\gamma$ mean free path for such gamma-rays is roughly
\begin{equation}
\lambda_{\gamma\gamma,pk} = \frac{1}{0.26 \sigma_T n_{\rm CMB,pk}}
\approx 4.6 \left( \frac{1+z}{10} \right)^{-3} {\rm pc},
\end{equation}
where $\sigma_T$ is the Thomson cross section.

\begin{figure}[t]
\begin{center}
\includegraphics[width=12cm]{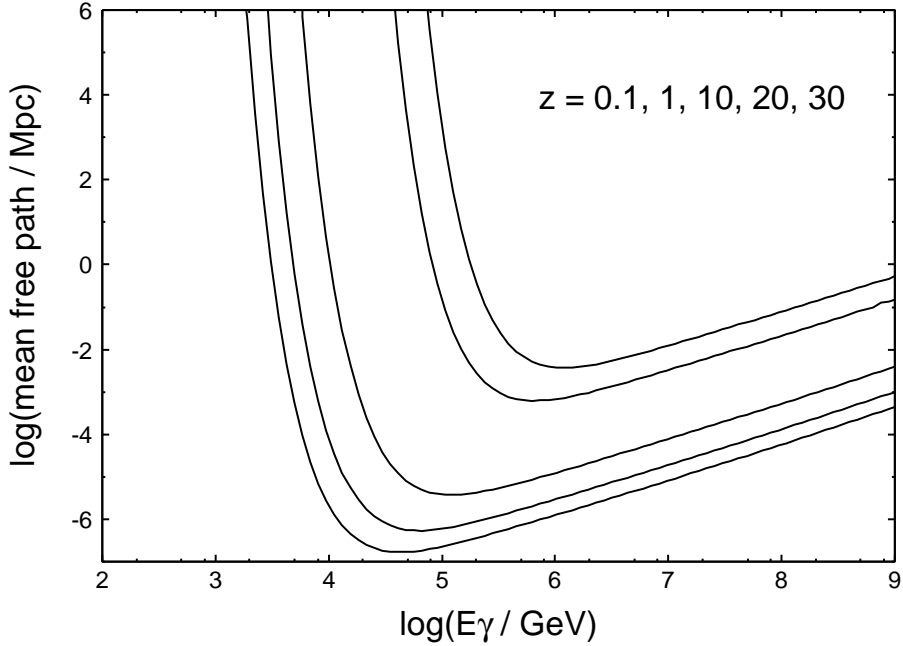}
\end{center}
\caption{$\gamma\gamma$ mean free path in the CMB vs. rest-frame gamma-ray energy $E'_\gamma$,
for $z=0.1,1, 5, 10, 20, 30$, from right to left.
%\Red{[Should include $z=5$ curve. Add "$\gamma\gamma$" before "mean free path" in Y-axis label.]}
\label{fig:mfp}}
\end{figure}

Proper evaluation of the $\gamma\gamma$ mean free path $\lambda_{\gamma\gamma}$
for arbitrary gamma-ray energies requires a convolution of
the energy- and angle-dependent pair production cross section $\sigma_{\gamma\gamma}$
\citep[e.g.][]{Berestetsky:1982aq} over the CMB spectrum, and is plotted for selected redshifts in Fig. \ref{fig:mfp}.
%in terms of the rest-frame gamma-ray energy
%\Cyan{[Must specify "rest-frame" energy!]}
For $E'_\gamma \lesssim E'_{\gamma,pk}$,
$\lambda_{\gamma\gamma}(E'_\gamma)$ is determined by the density of CMB photons
whose energies are $\epsilon'_{\rm CMB} \sim m_e^2/E'_\gamma$,
corresponding to the peak of $\sigma_{\gamma\gamma}$
and reflecting the Wien shape of the CMB spectrum.
In contrast, for $E'_\gamma \gtrsim E'_{\gamma,pk}$,
only the CMB photons with $\epsilon'_{\rm CMB} \sim \epsilon'_{\rm CMB,pk}$ are relevant,
and the shape of $\lambda_{\gamma\gamma}(E'_\gamma)$ is due to the high-energy tail of $\sigma_{\gamma\gamma}$.
%\Red{[As Nakamura-san also pointed out, this kind of explanation is necessary and helpful.]}

%%%%%%%%%%%%%%%%%%%%%%%%%%%%%%%%%%%%%%%%%%%%%%%%%%%%%%%%%%%%%%%%%%%
%%%%%%%%%%%%%%%%%%%%%%%%%%%%%%%%%%%%%%%%%%%%%%%%%%%%%%%%%%%%%%%%%%%

Assuming that the source spectrum extends to sufficiently high energies
for $\gamma\gamma$ interactions with the CMB (see \S \ref{sec:discussion}),
primary gamma-rays with energy $E'_\gamma$ would produce
electron-positron pairs with energies $E_e \approx E'_\gamma/2$.
These upscatter ambient CMB photons to generate a pair echo with average energy
\begin{equation}
\bar{E}_{\rm echo}^{\prime} %\langle E_{\rm echo}^{\prime} \rangle
= \epsilon'_{\rm CMB} \left(\frac{E_e}{m_e}\right)^2 %2.7 T_{\rm CMB} \frac{E_e^2}{m_e^2}
= 16 \left( \frac{1+z}{10} \right)^{-1}
  \left( \frac{E'_{\gamma}}{E'_{\gamma,pk}(z)} \right)^2
  {\rm TeV},
\label{E_pa_prime}
\end{equation}
%\Red{[Changed suffix explicitly to "echo" for clarity. Also, what is the factor 2.7 in the original equation?
%I don't understand, so I haven't changed the numerical values.]}
which would be observed at $z=0$ with energy
\begin{equation}
\bar{E}_{\rm echo} %\langle E_{\rm pa} \rangle
= \frac{\bar{E}_{\rm echo}^{\prime}}{1+z}
= 1.6 \left( \frac{1+z}{10} \right)^{-2}
  \left( \frac{E'_{\gamma}}{E'_{\gamma,pk}(z)} \right)^2
  {\rm TeV}.
\label{E_pa}
\end{equation}
Considering only a narrow range of $E'_\gamma$,
the corresponding echo spectrum will have a turnover above $\bar{E}'_{\rm echo}$,
as well as a power-law tail below $\bar{E}'_{\rm echo}$
with photon index $\sim 1.5$ from pairs undergoing IC cooling.
The total echo spectrum will be a superposition of such spectra over the range of $E'_\gamma$
that is effectively absorbed via $\gamma\gamma$ interactions (\S \ref{sec:results}).
The mean free path $\lambda_{\rm IC}$ and cooling length $\Lambda_{\rm IC}$
of the pairs for IC scattering with CMB peak photons are respectively
\begin{eqnarray}
\lambda_{\rm IC,pk}
&=& \frac{1}{\sigma_T n_{\rm CMB,pk}}
\approx 1.2 \left( \frac{1+z}{10} \right)^{-3} {\rm pc},
\\
\Lambda_{\rm IC,pk}
&=& \frac{3 m_e^2}{4 E_e \sigma_T U_{\rm CMB}}
\approx
  1.2 \left( \frac{1+z}{10} \right)^{-3}
  \left( \frac{E'_{\gamma}}{E'_{\gamma,pk}(z)} \right)^{-1}
  {\rm pc},
\label{lambda_IC_cool}
\end{eqnarray}
where $U_{\rm CMB}$ is the energy density of the CMB.
%\Red{[Added "pk" to suffix, and changed symbols for clarity.]}

The time delay between the arrival of the primary photons and the secondary pair echo
are caused by two effects.
The first is due to the intrinsic angular spread in the $\gamma\gamma$ and IC processes,
which is unavoidable even in the absence of magnetic fields.
The characteristic delay time in the observer frame from angular spreading is
\begin{equation}
\Delta t_{\rm A}
= \frac{1+z}{2 (E_e/m_e)^2}
  (\lambda_{\gamma\gamma} + \Lambda_{\rm IC})
\approx
  0.95 \times 10^{-6}~{\rm sec}
  \left( \frac{E'_\gamma}{E'_{\gamma,pk}(z)} \right)^{-2}
  \left( \frac{\lambda_{\gamma\gamma}}{\lambda_{\gamma\gamma,pk}}
  \right),
\end{equation}
where we have assumed $\lambda_{\gamma\gamma} \gg \Lambda_{\rm IC}$ (see below for justification).
The second effect, of our main interest here, is due to deflections of the pairs by magnetic fields,
whose characteristic delay time is
\begin{equation}
\Delta t_B
= \frac{1+z}{2}
  (\lambda_{\gamma\gamma} + \Lambda_{\rm IC})
  \theta_B^2
\approx
  3.8~{\rm sec}
  \left( \frac{1+z}{10} \right)^{-6}
  \left( \frac{E'_{\gamma}}{E'_{\gamma,pk}(z)} \right)^{-4}
  \left( \frac{B}{10^{-12}~{\rm G}} \right)^2
  \left( \frac{\lambda_{\gamma\gamma}}{\lambda_{\gamma\gamma,pk}} \right),
\label{t_B}
\end{equation}
where
\begin{equation}
\theta_B
= \frac{\Lambda_{\rm IC}}{r_{\rm L}}
\approx
  5 \times 10^{-5}
  \left( \frac{1+z}{10} \right)^{-2}
  \left( \frac{E'_{\gamma}}{E'_{\gamma,pk}(z)} \right)^{-2}
  \left( \frac{B}{10^{-12}~{\rm G}} \right),
\label{theta_B}
\end{equation}
is the average deflection angle of the pairs when the fields are coherent over scales of $\Lambda_{\rm IC}$,
\begin{equation}
r_{\rm L} = \frac{E_e}{eB}
\approx
  30 \left( \frac{1+z}{10} \right)^{-1}
  \left( \frac{E'_\gamma}{E'_{\gamma,pk}(z)} \right)
  \left( \frac{B}{10^{-12}~{\rm G}} \right)^{-1}
  {\rm kpc},
\end{equation}
is the Larmor radius of the pairs, and $B$ is the comoving amplitude of the magnetic field,
which is related to the physical amplitude of the magnetic field $B'$ in the rest-frame at $z$ as $B=B' (1+z)^{-2}$,
following the convention in the literature on IGMFs.
%\Red{[This description OK?]}
Eqs. \ref{t_B}-\ref{theta_B} can also be straightforwardly adapted to the case of fields
randomly tangled on scales smaller than $\Lambda_{\rm IC}$ \citep{2008ApJ...682..127I}.
The ratio of the two delay timescales are
\begin{equation}
\frac{\Delta t_{\rm A}}{\Delta t_B}
\approx 
  2.5 \times 10^{-7}
  \left( \frac{1+z}{10} \right)^6
  \left( \frac{E'_{\gamma}}{E'_{\gamma,pk}(z)} \right)^2
  \left( \frac{B}{10^{-12}~{\rm G}} \right)^{-2}.
\label{tA_tB}
\end{equation}

At face value, $\Delta t_B$ for pair echos from a GRB at $z \approx 10$
would be in the observationally interesting range of several to tens of seconds,
as long as the ambient magnetic fields are of order $B \approx 10^{-12}~{\rm G}$
at distances of $\lambda_{\gamma\gamma} \approx 5~{\rm pc}$ from the GRB.
This is interestingly close to some recent predictions from
numerical simulations of magnetic field generation in Pop III star forming regions \citep{2008ApJ...688L..57X}.
%(\S \ref{sec:discussion}).
%It is interesting to compare these typical numbers with those
%from numerical simulations of magnetogenesis at population-III
%star formation \citep{2008ApJ...688L..57X}. According to it,
%magnetic fields of order $10^{-13.5} \sim 10^{-12.5}~{\rm G}$
%are generated at distances $1 \sim 10~{\rm pc}$ from the center
%in a star forming halo.
However, for echo photons resulting from primary gamma-rays with $E'_\gamma \sim E'_{\gamma,pk}$,
Eq. (\ref{E_pa_prime}) shows that $\bar{E}'_{\rm echo}$ would still be so high
that most of them are absorbed locally by further $\gamma\gamma$ interactions with the CMB
on scales $\lambda_{\gamma\gamma,{\rm echo}} \sim 1~{\rm kpc}$ at $z = 10$.
Thus, we focus on the low-energy portion of the echo spectrum
unaffected by secondary $\gamma\gamma$ absorption,
which arise mainly from primary gamma-rays with energies sufficiently lower than $E'_{\gamma,pk}$
interacting with the CMB Wien regime where
$\lambda_{\gamma\gamma} \gg \lambda_{\gamma\gamma,pk}$ (Fig. \ref{fig:mfp}).
The relevant delay time can then be substantially longer for the same $B$,
or conversely, much weaker $B$ can be probed on the same timescales (Eq. \ref{t_B}).
The weakest field strengths that can be probed through such pair echos
is $B \sim 10^{-16}~{\rm G}$ for $z = 10$,
determined by the condition that $\Delta t_B = \Delta t_A$ (Eq. \ref{tA_tB}).

Thus, the unabsorbed part of high-$z$, CMB-triggered pair echos
allows us to probe magnetic fields with amplitudes $B \gtrsim 10^{-16}~{\rm G}$,
at distances $\lambda_{\gamma\gamma} \gtrsim 10~{\rm kpc}$ from the GRB.
On these scales, the relevant magnetic fields should be associated with the IGM,
since the collapsed halos within which GRB occur are likely to be smaller than present-day galaxies
at $z \gtrsim 5-10$ \citep{2001PhR...349..125B}.
IGMFs of order $B \sim 10^{-16}~{\rm G}$ have been predicted by some models involving
cosmic reionization fronts \citep{2000ApJ...539..505G,2005A&A...443..367L},
%as well as some cosmological models
for which high-$z$ GRB pair echos may provide a valuable probe.

%2006Sci...311..827I.

%\Red{["At this distance scale, magnetic fields generated cosmologically
%may be stronger than those produced in the star forming halo" is highly unlikely,
%because the halo collapses out of the IGM with the cosmological fields.
%Ichiki et al. 2006 reference removed here, OK?]}

\subsection{Numerical formulation \label{sec:formulation}}

Here we briefly summarize our formulation for numerical calculations of
the spectra and light curves of GRB pair echos.
%for results in \S \ref{sec:results}.
%of GRB pair echos can be evaluated as follows.
For a GRB with primary fluence
$dN_{\gamma}/dE_{\gamma}$, the time-integrated flux of secondary
pairs during the GRB duration is
\begin{equation}
\frac{dN_{e,{\rm 0}}}{d\gamma_e} (\gamma_e)
= 4 m_e
  \frac{dN_{\gamma}}{dE_{\rm \gamma}}(E_{\gamma} = 2 m_e \gamma_e)
  \left[ 1 - e^{-\tau_{\gamma\gamma}(E_{\gamma} = 2 \gamma_e m_e)}
  \right],
\label{eq:dN0dgamma}
\end{equation}
where $\tau_{\gamma \gamma}(E_\gamma)$ is the optical depth to
$\gamma-\gamma$ pair production for gamma-rays with energy
$E_\gamma$. The time-dependent spectrum of the pair echo is
\begin{equation}
\frac{d^2 N_{\rm echo}}{dt dE_\gamma}
= \int d\gamma_e \frac{dN_e}{d{\gamma_e}}
  \frac{d^2 N_{\rm IC}}{dt dE_\gamma},
\end{equation}
where $d^2 N_{\rm IC}/dt dE_\gamma$ is the IC power from
a single electron or positron, and $dN_e/d{\gamma_e}$ is
the total time-integrated flux of pairs responsible for
the echo emission observed at time $t_{\rm obs}$ after the burst,
which is related nontrivially to $dN_{e, \rm 0}/d{\gamma_e}$ in Eq. (\ref{eq:dN0dgamma}).
This expression was evaluated by \citet{2008ApJ...682..127I} 
taking into proper account
the relevant geometrical effects and the stochastic nature of magnetic deflections.
Although numerical integration is required to obtain the end results, it can be roughly
approximated by
$dN_e/d{\gamma_e} =
 (\lambda_{\rm IC, cool}/c \Delta t) dN_{e,{\rm 0}}/d\gamma_e$
\citep{2002ApJ...580L...7D}. 

\section{Results \label{sec:results}}
%%%%%%%%%%%%%%%%%%%%%%%%%%%%%%%%%%%%%%%%%%%%%%%%%%%%%%%%%%%%%%%%%%%

Our numerical results employing the formulation of \S \ref{sec:formulation} are presented below.
Regarding the properties of the GRB primary emission,
we assume a constant spectrum
%\begin{equation}
%\frac{dN_\gamma}{dE_\gamma} \propto E_\gamma^{-2},%
%\end{equation}
$dN_\gamma/dE_\gamma \propto E_\gamma^{-2}$ for $1~{\rm TeV} \leq E_\gamma \leq 100~{\rm TeV}$
during a duration $t_{\rm GRB} = 10~{\rm sec}$.
Note that the GRB spectrum from the early afterglow 
may quite plausibly extend up to $\sim$1-10 TeV during the first $\sim$10-100 sec
\citep{2010ApJ...712.1232W}, although it remains to be seen
whether this holds up to 100 TeV \citep[see also][]{2004ApJ...613.1072R}.
We also take an isotropic-equivalent total energy $E_{\rm tot} = 10^{55}~{\rm erg}$,
corresponding to the most luminous GRBs observed so far \citep{2009Sci...323.1688A}.

\begin{figure}
\begin{center}
\includegraphics[width=12cm]{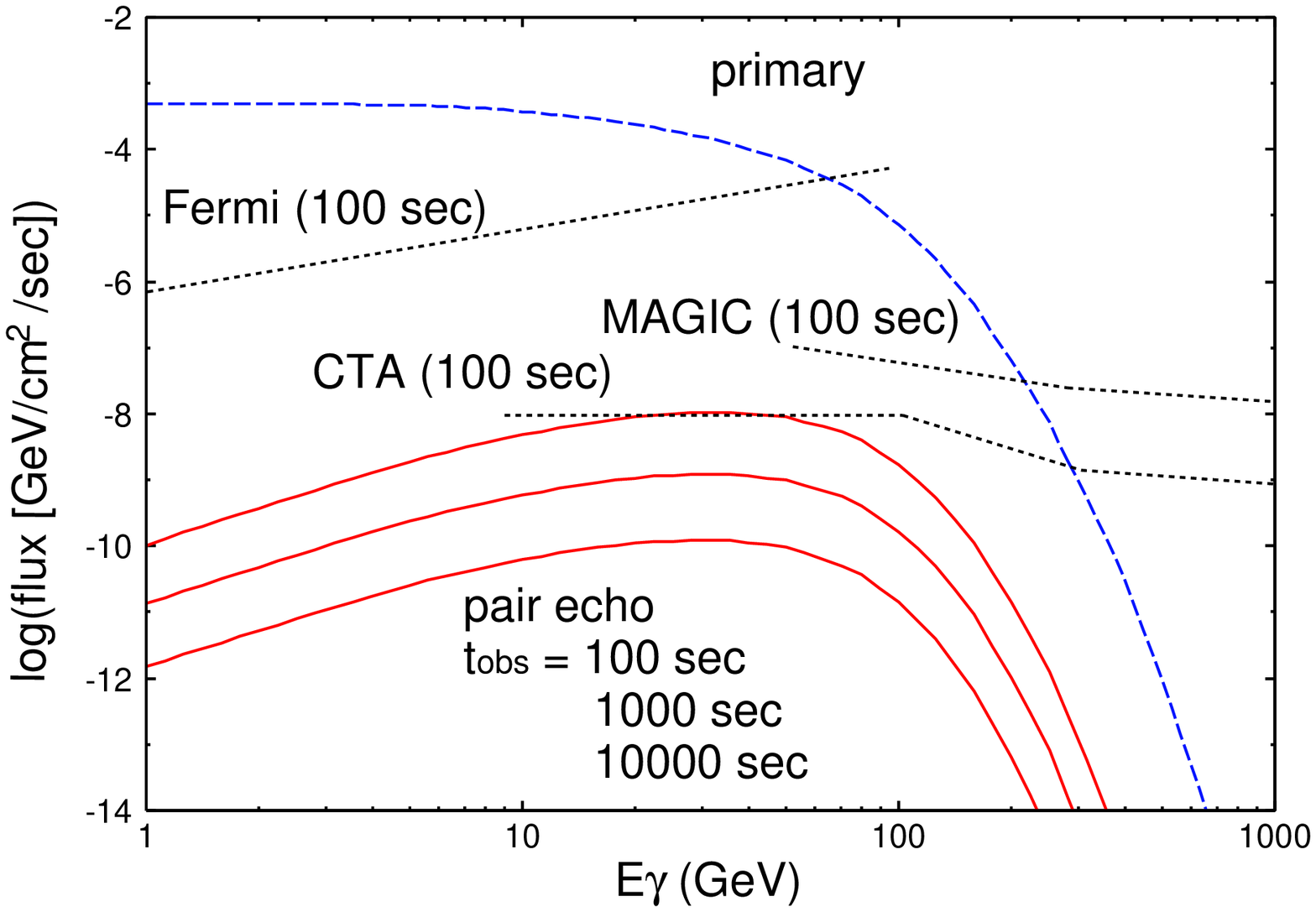}
\end{center}
\caption{Spectra of the primary emission (dashed curve) and those of the pair echo (solid curves)
at observer times $t_{\rm obs} = 10^2, 10^3, 10^4~{\rm sec}$, from top to bottom,
for $z = 10$, $B = 10^{-15}~{\rm G}$ and the low-EBL case.
Overlayed are $5-\sigma$ sensitivities for Fermi, MAGIC and CTA for integration times of $100~{\rm sec}$.
\label{fig:spectrum_low-CIB}}
\hspace{1cm}
\begin{center}
\includegraphics[width=12cm]{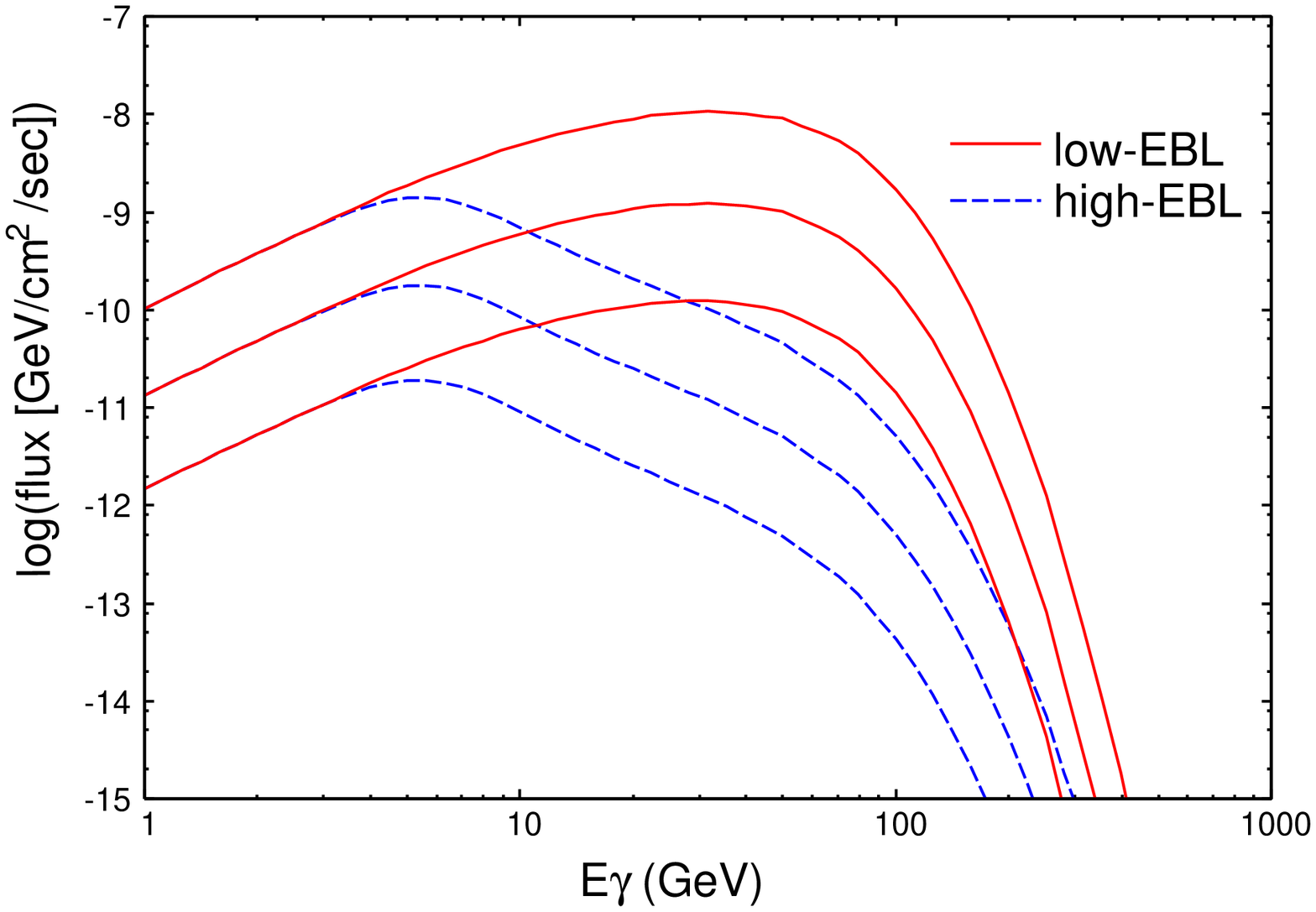}
\end{center}
\caption{Comparison of pair-echo spectra between the low-EBL and high-EBL cases,
for $z = 10$, $B = 10^{-15}~{\rm G}$ and $t_{\rm obs} = 10^2, 10^3, 10^4~{\rm sec}$, from top to bottom.
\label{fig:CIB-comparison}}
\end{figure}

Fig. \ref{fig:spectrum_low-CIB} shows the spectra of the primary emission
together with those of the pair echo at observer times $t_{\rm obs} = 10^2, 10^3, 10^4~{\rm sec}$,
for $z = 10$, $B = 10^{-15}~{\rm G}$, and the low-EBL case.
Due to absorption by the EBL at low $z$ (Fig.\ref{fig:tau_comp}),
both the primary and pair echo emission are substantially
attenuated at $E_\gamma \gtrsim 100~{\rm GeV}$.
To be compared are estimated $5-\sigma$ detection sensitivities for the Fermi\footnote{http://fermi.gsfc.nasa.gov/},
MAGIC\footnote{http://wwwmagic.mppmu.mpg.de/},
and CTA\footnote{http://www.cta-observatory.org/} telescopes, for exposure times of $100~{\rm sec}$.
Although far from the capabilities of current instruments such as Fermi or MAGIC,
the pair echo at $t_{\rm obs} = 100~{\rm sec}$ may be marginally detectable by the next-generation facility CTA,
or similar projects such as AGIS\footnote{http://www.agis-observatory.org/} or 5@5 \citep{2001APh....15..335A}.
Note that the slewing time of the large size telescopes of CTA are projected to be comparable to MAGIC,
i.e. 180 deg in $\sim$20 sec.
Detection at later times would be more difficult
as the pair echo flux decreases as $\sim t^{-1}$ while the sensitivities scale as $t^{-1/2}$,

In Fig. \ref{fig:CIB-comparison}, we compare the pair-echo spectra for the low-EBL and high-EBL cases,
maintaining $z = 10$ and $B = 10^{-15}~{\rm G}$.
As apparent in Fig. \ref{fig:tau_comp}, the differences between the two high-$z$ EBL models
are most significant at $10~{\rm GeV} \lesssim E_\gamma \lesssim 100~{\rm GeV}$.
In particular, the spectral peaks at $\sim 40~{\rm GeV}$ noticeable in the low-EBL case
are dramatically obliterated in the high-EBL case, considerably reducing their observability.
Thus, we concentrate on the low-EBL case below.

%%%%%%%%%%%%%%%%%%%%%%%%%%%%%%%%%%%%%%%%%%%%%%%%%%%%%%%%%%%%%%%%%%%
%%%%%%%%%%%%%%%%%%%%%%%%%%%%%%%%%%%%%%%%%%%%%%%%%%%%%%%%%%%%%%%%%%%

\begin{figure}
\begin{center}
\includegraphics[width=12cm]{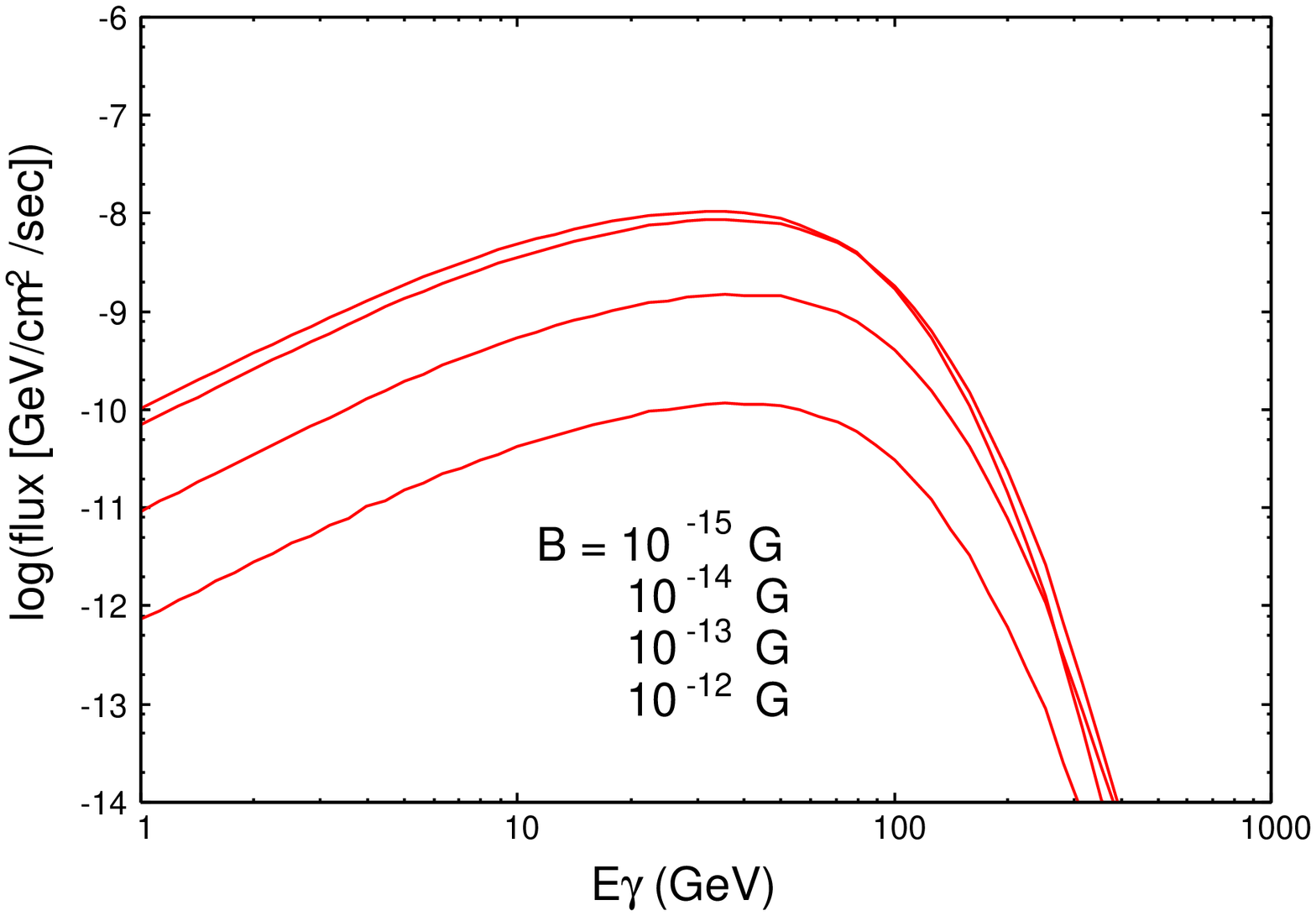}
\end{center}
\caption{Pair-echo spectra for $z = 10$, $t_{\rm obs} = 100~{\rm sec}$, the low-EBL case,
and $B=10^{-15},10^{-14},10^{-13}$ and $10^{-12}~{\rm G}$, from top to bottom.
\label{fig:B-dependence}}
\hspace{1cm}
\begin{center}
\includegraphics[width=12cm]{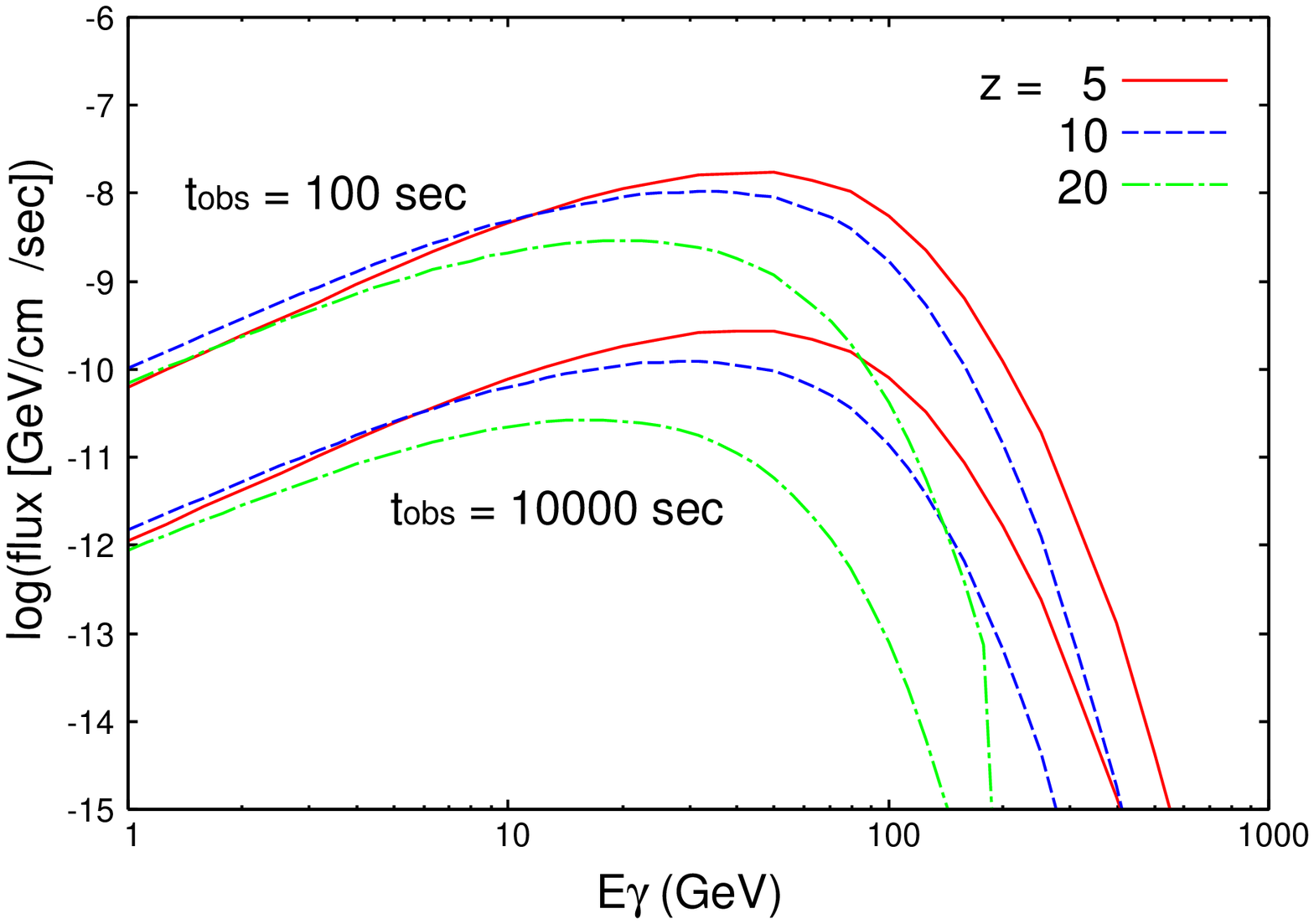}
\end{center}
\caption{Pair-echo spectra for GRB redshifts $z=$5 (solid),10 (dashed) and 20 (dot-dashed),
compared at $t_{\rm obs} = 100$ (top) and $10^4~{\rm sec}$ (bottom),
for $B = 10^{-15}~{\rm G}$ and the low-CIB case.
\label{fig:z-dependence}}
\end{figure}

The dependence of the pair-echo spectra on the magnetic field amplitude $B$
is shown in Fig. \ref{fig:B-dependence}, for $z = 10$, $t_{\rm obs} = 100~{\rm sec}$, and the low-EBL case.
As can be seen from Eq. (\ref{t_B}),
higher-energy primary gamma-rays contribute more to the pair echo when compared at fixed $t_{\rm obs}$,
leading to larger average energies of the pair echo (Eq. (\ref{E_pa})).
Since the echo at $E_\gamma \gtrsim 100~{\rm GeV}$ is largely absorbed by the EBL,
the observed flux is lower for stronger fields, as long as $B \gtrsim 10^{-16}~{\rm G}$.
However, for $B \lesssim 10^{-16}~{\rm G}$,
the delay timescale becomes dominated by angular spreading (Eq. \ref{tA_tB}),
and the pair echo properties become independent of $B$.

Fig. \ref{fig:z-dependence} compares the pair-echo spectra for different GRB redshifts,
at fixed observer times $t_{\rm obs} = 100~{\rm sec}$ and $10^4~{\rm sec}$,
$B = 10^{-15}~{\rm G}$ and the low-EBL case.
Besides the obvious trend of the pair echo being fainter for higher $z$, 
a sharp cut off due to the absorption by CMB can be seen
at the highest energies for $z=20$.

%%%%%%%%%%%%%%%%%%%%%%%%%%%%%%%%%%%%%%%%%%%%%%%%%%%%%%%%%%%%%%%%%%%
%%%%%%%%%%%%%%%%%%%%%%%%%%%%%%%%%%%%%%%%%%%%%%%%%%%%%%%%%%%%%%%%%%%

%%%%%%%%%%%%%%%%%%%%%%%%%%%%%%%%%%%%%%%%%%%%%%%%%%%%%%%%%%%%%%%%%%%
\section{Discussion \label{sec:discussion}}
%%%%%%%%%%%%%%%%%%%%%%%%%%%%%%%%%%%%%%%%%%%%%%%%%%%%%%%%%%%%%%%%%%%

Despite their obviously lower fluxes and harder observability,
a prime advantage for considering pair echos from high-$z$ GRBs 
is that they probe ambient magnetic fields at epochs
that are much less polluted by magnetization from galactic winds or quasar outflows
whose activity peak at later times
\citep{2001ApJ...556..619F,2006MNRAS.370..319B}.
They should therefore be more sensitive to magnetic field generation processes in the early universe,
either during the cosmic reionization era or even earlier epochs.
The observationally favorable field amplitudes of $B \sim 10^{-16}$-$10^{-15}$ G
that we find is in the range predicted by the Biermann battery mechanism \citep{2000ApJ...539..505G}
or radiation drag effects at cosmic reionization fronts \citep{2005A&A...443..367L}
(see however, \citet{2010ApJ...716.1566A}). 
Some cosmological mechanisms may even result in IGMFs of such strengths \citep{2008PhRvL.101q1302C}.
Note also that this is close to the claimed IGMF strengths deduced from some recent analyses
of {\it Fermi} data on blazars at lower $z$ \citep{2010Sci...328...73N,2010arXiv1005.1924A},
so high-$z$ GRBs may provide an independent test of their existence and origin.

%Pop III
%It is interesting to compare these typical numbers with those
%from numerical simulations of magnetogenesis at population-III
%star formation \citep{2008ApJ...688L..57X}. According to it,
%magnetic fields of order $10^{-13.5} \sim 10^{-12.5}~{\rm G}$
%are generated at distances $1 \sim 10~{\rm pc}$ from the center
%in a star forming halo.

%There are both advantage and disadvantage of considering
%high-z GRBs compared with low-z GRBs. An obvious disadvantage
%is the large luminosity distances of high-z GRBs.
%In fact, \cite{2008ApJ...687L...5T} showed that pair echos
%from GRBs with standard luminosity would be observable only
%when the redshift is relatively small, $z \lesssim 1$.
%Therefore, we expect that pair echos only from very energetic
%GRBs would be observable for $z \gtrsim 10$.

%\Cyan{[NOTE: Changed order of advantages]}
%On the other hand, there are a couple of advantages.

A further point to mention is that the intergalactic radiation field relevant
for the primary $\gamma\gamma$ interaction may be dominated by the well-understood CMB,
in contrast to lower $z$ where the corresponding EBL is relatively uncertain.
However, this does require the primary GRB spectrum to extend up to very high, multi-TeV energies,
which is not guaranteed at the moment.
Furthermore, we have seen that the high-energy end of the secondary pair-echo gamma-rays
can still be significantly affected by the high-$z$ EBL.
In this regard, a somewhat different type of pair echo emission can result from
primary $\gamma\gamma$ interactions with the high-$z$ EBL.
Although not discussed here, this may also be worth consideration
as the necessary primary photon energies are in the much more modest range of $\lesssim$TeV.

%GRB spectrum extension to multi-TeV.

%%%%%%%%%%%%%%%%%%%%%%%%%%%%%%%%%%%%%%%%%%%%%%%%%%%%%%%%%%%%%%%%%%%
\section{Summary \label{sec:summary}}
%%%%%%%%%%%%%%%%%%%%%%%%%%%%%%%%%%%%%%%%%%%%%%%%%%%%%%%%%%%%%%%%%%%

In this paper, we have studied the expected properties of pair echos from high-z GRBs,
their detectability, and the consequent implications for probing the IGMF.
At $z \gtrsim 5$, the CMB may constitute the most relevant intergalactic radiation field
for the primary $\gamma\gamma$ interaction.
We found that pair echos from luminous GRBs at $z \lesssim 10$
may be observable with next generation gamma-ray telescopes such as CTA
as long as the primary GRB spectra extend to multi-TeV energies,
the IGMF strengths are $B \sim 10^{-16}-10^{-15}~{\rm Gauss}$,
and the EBL is relatively low.
Although their actual detection may be quite challenging,
they would provide us with a unique way to probe IGMFs at early epochs
thay may have originated during the cosmic reionization era and beyond.

\acknowledgments

This work is supported in part by the Grant-in-Aid from the 
Ministry of Education, Culture, Sports, Science and Technology
(MEXT) of Japan, No.19540283, No.19047004(TN), No.21840028(KT),
and No.22540278(SI), and by the Grant-in-Aid for the global COE program
{\it The Next Generation of Physics, Spun from Universality and Emergence}
at Kyoto University and "Quest for Fundamental Principles in
the Universe: from Particles to the Solar System and the Cosmos"
at Nagoya University from MEXT of Japan.

\bibliographystyle{hapj}
\bibliography{grb}

\end{document}